\documentclass[10pt,conference]{IEEEtran}

\usepackage{amsmath,amssymb,amsfonts}
\usepackage{algorithmic}
\usepackage{graphicx}
\usepackage{textcomp}
\def\BibTeX{{\rm B\kern-.05em{\sc i\kern-.025em b}\kern-.08em
    T\kern-.1667em\lower.7ex\hbox{E}\kern-.125emX}}
    
\PassOptionsToPackage{table,xcdraw}{xcolor}

\AtBeginDocument{%
  \providecommand\BibTeX{{%
    Bib\TeX}}}

\usepackage[english]{babel}
\usepackage[utf8]{inputenc}

\usepackage{flushend}
\usepackage{balance}
\usepackage{multirow}

\usepackage{color}
\usepackage{booktabs}
\usepackage{fancybox}
\usepackage{float}
\usepackage{array,graphicx}
\usepackage[flushleft]{threeparttable}
\usepackage[most]{tcolorbox}
\usepackage{xspace}
\usepackage{listings}
\usepackage{pifont}
\usepackage{tikz}
\usetikzlibrary{shapes.geometric}
\usepackage{blindtext}
\usepackage{colortbl}

\usepackage{makecell}
\usepackage{stfloats}
\usepackage{url}

\usepackage{colortbl} 

\usepackage{booktabs}

\usepackage{hyperref}

\usepackage{tcolorbox}

\usepackage{comment}

\definecolor{purplish}{HTML}{D8D0E3}
\definecolor{purplishlight}{HTML}{EBE7F1}
\definecolor{purplishdark}{HTML}{875afb}

\newtcolorbox[auto counter,number within=section]{rqbox}[2]{
    nameref=#1,
    title=\small{#1}, 
    enhanced,
    attach boxed title to top left={yshift=-6pt, xshift=8pt},
    boxed title style={size=small,boxsep=1pt},
    colframe=purplishdark,colback=white,colbacktitle=purplishdark,
    boxsep=2pt,left=2pt,right=2pt,top=6pt,bottom=2pt,middle=2pt
}

\begin{document}

\title{Great Power Brings Great Responsibility: Personalizing Conversational AI for Diverse Problem-Solvers}


\author{\IEEEauthorblockN{Italo Santos\IEEEauthorrefmark{1}, Katia Romero Felizardo\IEEEauthorrefmark{2}, Igor Steinmacher\IEEEauthorrefmark{1} and Marco A. Gerosa\IEEEauthorrefmark{1}}
\IEEEauthorblockA{\IEEEauthorrefmark{1}Northern Arizona University, Flagstaff, AZ, USA\\}
\IEEEauthorblockA{\IEEEauthorrefmark{2}Federal Technological University of Paraná, PR, Brazil\\}
Email: italo.santos@nau.edu, katiascannavino@utfpr.edu.br, \\igor.steinmacher@nau.edu, marco.gerosa@nau.edu}

\maketitle
\IEEEpeerreviewmaketitle

\begin{abstract}
    Newcomers onboarding to Open Source Software (OSS) projects face many challenges. Large Language Models (LLMs), like ChatGPT, have emerged as potential resources for answering questions and providing guidance, with many developers now turning to ChatGPT over traditional Q\&A sites like Stack Overflow.
    Nonetheless, LLMs may carry biases in presenting information, which can be especially impactful for newcomers whose problem-solving styles may not be broadly represented. This raises important questions about the accessibility of AI-driven support for newcomers to OSS projects.
    This vision paper outlines the potential of adapting AI responses to various problem-solving styles to avoid privileging a particular subgroup.
    We discuss the potential of AI persona-based prompt engineering as a strategy for interacting with AI.   
    This study invites further research to refine AI-based tools to better support contributions to OSS projects.
\end{abstract}


\section{Introduction}
\label{sec:introduction}


Newcomers often face significant barriers when beginning to contribute to Open Source Software (OSS) projects~\cite{steinmacher2015social, pinto2017training, pinto2019training, santos2022hits}. Large Language Models (LLMs), like ChatGPT, have emerged as potential resources for addressing these onboarding challenges, with many users now turning to ChatGPT over traditional Q\&A sites like Stack Overflow~\cite{kabir2024stack, StackOverflow2023}. Yet, because LLMs are trained on existing data, they may inadvertently carry biases when presenting information. This raises important questions about the accessibility of AI-driven support for newcomers to OSS projects.

Research highlights that information presentation in OSS projects~\cite{padala2020gender, mendez2018open}---such as documentation and issue descriptions---often favors specific problem-solving styles (e.g., hands-on learners) over others (e.g., process-oriented learners), contributing to gender biases. While additional studies have further examined gender bias in OSS~\cite{murphy2024gendermag, ford2017someone, ford2016paradise, nafus2012patches, vasilescu2015gender}, there is still a need to adapt OSS environments to support diverse contributors effectively~\cite{santosvlhcc}. The rise of generative conversational artificial intelligence (AI), like ChatGPT, which reached over 100 million users in two months~\cite{hu2023chatgpt}, introduces new opportunities to address this gap by providing personalized, style-aligned support to diverse contributors.

Skjuve et al.~\cite{skjuve2024people} reported that conversational AI is perceived as highly capable of handling cognitive tasks that were once considered difficult to automate~\cite{frey2017future}. The decline in Stack Overflow traffic is attributed to ChatGPT's rise~\cite{kabir2024stack, StackOverflow2023}. Moreover, people increasingly rely on ChatGPT to simplify complex concepts they need to learn~\cite{sundar2020rise}. ChatGPT's potential to be adapted to address bias by delivering responses tailored to individual problem-solving styles offers a promising way to lower entry barriers for underrepresented groups in OSS. 
ChatGPT is a good fit for our study due to its widespread adoption, training on diverse knowledge sources, and ability to generate context-aware responses~\cite{hu2023chatgpt}. 


This paper offers a conceptual exploration to foster discussions and reflection on how AI might empower newcomers to contribute to OSS projects, considering their different problem-solving styles. In this vision paper, we seek to examine the possibilities, outline future directions, and call the research community to establish best practices that balance personalized AI assistance with skill development and personal growth in OSS environments.




\section{Related work}
\label{sec:backrelated}

\textbf{Problem-solving styles:} Research reveals that developers exhibit diverse problem-solving styles~\cite{burnett2016gendermag} and motivations~\cite{gerosa2021shifting}. For instance, studies indicate that women are often task-oriented, whereas men are frequently motivated to explore new technologies for enjoyment~\cite{padala2020gender, mendez2018open, mendez2018gender}. These problem-solving differences impact how newcomers engage with OSS. In this vision paper, we would like to address how we can personalize LLMs' responses to help newcomers contribute to OSS projects.

\textbf{Diversity in OSS:} Low diversity within OSS communities is a well-documented concern across various dimensions, including gender~\cite{trinkenreich2021women, guizani2022debug, bosu2019diversity, terrell2016gender}, language~\cite{storey2016social}, and geographic location~\cite{storey2016social}. Research shows that diverse teams are generally more productive~\cite{vasilescu2015gender}, yet minorities often face significant barriers when trying to join and thrive within OSS communities~\cite{trinkenreich2021women}. 
This vision paper aims to reduce these biases by leveraging LLMs to support diverse problem-solving styles, fostering a more inclusive OSS environment.

\section{Leveraging LLMs to Support Newcomers Problem-Solving Styles in OSS}
\label{sec:contextexample}




A persona is an archetype representing a group of target users who share common behavioral characteristics~\cite{pruitt2010persona}. Persona-based prompt engineering is a strategic method for interacting with AI, where prompts are crafted to generate responses that address the traits, behaviors, and perspectives of defined personas~\cite{white2023prompt}. We propose to aid newcomers in contributing to OSS projects by tailoring interactions to their problem-solving styles. For example, we can leverage personas from the GenderMag theoretical framework~\cite{burnett2016gendermag}, which characterizes problem-solving styles that cluster by gender: attitude toward risk, computer self-efficacy, motivation, information processing, and technology learning style. 
GenderMag is designed to improve technology's inclusiveness for problem-solving tasks 
and it is widely used in the literature for adapting traditional user interfaces~\cite{burnett2016finding, burnett2018gendermag, mendez2018gender, padala2020gender, guizani2022debug, santos2023designing}.  
GenderMag uses Abi and Tim personas to represent opposite ends of the spectrum: Abi reflects values often preferred by women, while Tim embodies those typically favored by men. 

LLM enables the development of intelligent agents capable of emulating specific characters or roles in behavior and dialogue~\cite{wang2023rolellm}. Gerosa et al.~\cite{gerosa2024can} describe persona-based prompting as harnessing LLMs ability to emulate interactions with specific demographic or psychographic profiles, creating virtual subjects with consistent traits. By designing tailored prompts, we can guide the AI in generating responses to aid newcomers contributing to OSS projects by tailoring interactions to their unique problem-solving styles. 

\begin{figure}[!ht]
    \centering
    \includegraphics[width=0.48\textwidth]{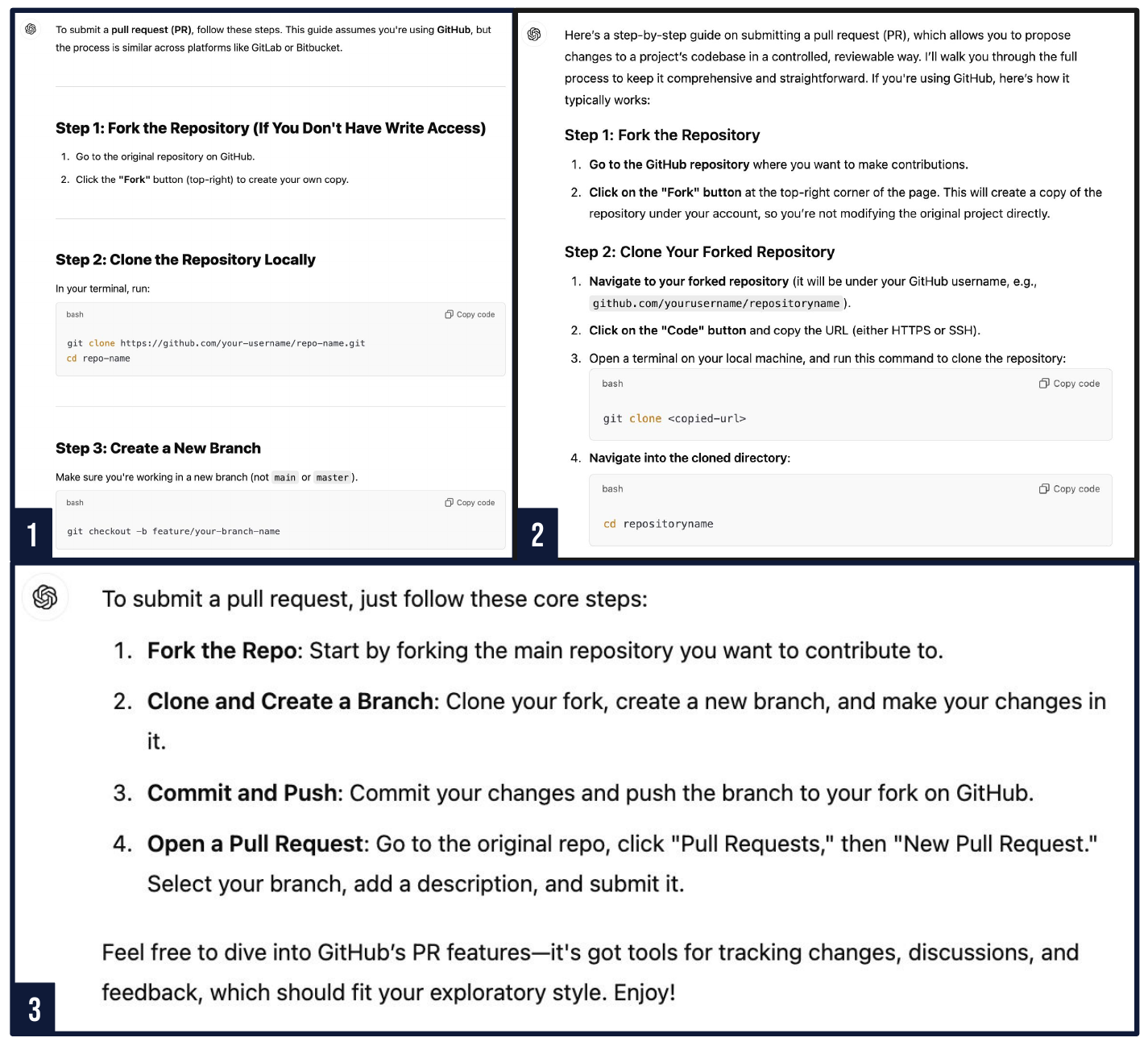}
    \vspace{-3mm}
    \caption{Using prompt engineering in a ChatGPT by asking, ``\textit{How can I submit a pull request?}" The conversation was generated using GPT-4. Panel (1) shows the ChatGPT response without persona-based prompt engineering, while Panels (2) and (3) display responses tailored to the GenderMag Abi and Tim personas, respectively.}
    \label{fig:exampleresponses}
\end{figure}

For this vision paper, we present the use case shown in Figure~\ref{fig:exampleresponses}. We created prompts guiding ChatGPT to answer the same query for personas representing GenderMag Abi and Tim: \textit{How can I submit a pull request?}. It is possible to compare the ChatGPT response without persona-based prompts in panel (1) with the tailored responses for the GenderMag personas Abi (2) and Tim (3). In Abi's response, the LLM outlines a structured, step-by-step process for submitting a pull request, aligning with Abi's process-oriented problem-solving style. It emphasizes the clarity and reversibility of each step, catering to Abi's risk-averseness. Conversely, for Tim's persona, the LLM provides a straightforward list of steps and suggests to ``tinker around'' after completion, reflecting Tim's exploratory and experimental problem-solving style. The replication package, including the chat transcripts, is publicly accessible~\cite{replicationpackage3}.

\section{Research opportunities}
\label{sec:researchopportunities}

Exploring AI to support diverse problem-solving styles in OSS contributions presents several research opportunities that align with our prior discussion.

\textbf{Empirical Studies Tailoring AI Responses:} Future research can investigate how variations in AI-generated guidance impact newcomers with different problem-solving styles. For example, structured instructions provided by the chatbot—such as step-by-step guidance—could benefit process-oriented learners who prefer clear directions, while exploratory learners might find open-ended prompts, which encourage independent thinking, more effective for exploring solutions. This research would analyze and refine AI-driven approaches tailored to these distinct preferences. Furthermore, future research can extend the use of the GenderMag framework to customize AI responses in OSS environments using different prompt engineering, fine-tuning, and other techniques. 

\textbf{Inferring Personas from Interaction:} Future research can develop methods for AI to infer a user's GenderMag persona from behavior rather than relying on questionnaires. Analyzing newcomers' preferences can enable real-time adaptation, enhancing the AI support. 

These opportunities drive research that refines AI tools, ensuring a more inclusive and supportive OSS environment for diverse contributors.

\section{Conclusion}
\label{sec:conclusion}

This paper discussed the potential of LLMs, like ChatGPT, to enhance newcomers' onboarding in OSS projects by tailoring responses to diverse problem-solving styles. 
Using persona-based prompt engineering, we illustrate the potential of LLMs to adapt to different problem-solving styles by generating outputs tailored to newcomers' unique characteristics. This approach highlights opportunities for LLMs in OSS and invites further research to refine AI-based tools to support contributions to OSS projects. 


\section*{Acknowledgments}
Katia Felizardo is funded by a research grant from the Brazilian National Council for Scientific and Technological Development (CNPq), Grant 302339/2022--1.


\bibliographystyle{IEEEtran}
\bibliography{IEEEabrv,bibtex.bib}

\end{document}